\documentclass{Interspeech2024}
\usepackage{hyperref}
\usepackage{tabularray}
\usepackage{adjustbox}
\usepackage{graphicx}
\usepackage{multirow}
\usepackage{hyperref}
\usepackage{url}
\usepackage{cite}
\usepackage{amsmath}

\interspeechcameraready

\title{WenetSpeech4TTS: A 12,800-hour Mandarin TTS Corpus for Large Speech Generation Model Benchmark}

\name[affiliation={1,\dagger}]{Linhan}{Ma}
\name[affiliation={1,\dagger}]{Dake}{Guo}
\name[affiliation={1}]{Kun}{Song}
\name[affiliation={1}]{Yuepeng}{Jiang}
\name[affiliation={2,3,5}]{Shuai}{Wang}
\name[affiliation={3}]{Liumeng}{Xue}
\name[affiliation={1}]{Weiming}{Xu}
\name[affiliation={1}]{Huan}{Zhao}
\name[affiliation={4,5}]{Binbin}{Zhang}
\name[affiliation={1,*}]{Lei}{Xie}

\address{
  $^1$Audio, Speech and Language Processing Group (ASLP@NPU), \\ School of Computer Science, Northwestern Polytechnical University, Xi'an, China\\
  $^2$Shenzhen Research Institute of Big Data, $^3$School of Data Science, \\ The Chinese University of Hong Kong, Shenzhen (CUHK-Shenzhen), China\\ 
  $^4$Shanghai Bigmelon Technology,   $^5$WeNet Open Source Community, China}
\email{\{mlh2023, guodake\}@mail.nwpu.edu.cn, lxie@nwpu.edu.cn\thanks{† Equal contribution. Random order.}\thanks{* Corresponding author.}}


\begin{document}

\maketitle

\begin{abstract}

    With the development of large text-to-speech (TTS) models and scale-up of the training data, state-of-the-art TTS systems have achieved impressive performance. In this paper, we present \textit{WenetSpeech4TTS}, a multi-domain Mandarin corpus derived from the open-sourced WenetSpeech dataset. Tailored for the text-to-speech tasks, we refined WenetSpeech by adjusting segment boundaries, enhancing the audio quality, and eliminating speaker mixing within each segment. Following a more accurate transcription process and quality-based data filtering process, the obtained WenetSpeech4TTS corpus contains $12,800$ hours of paired audio-text data. Furthermore, we have created subsets of varying sizes, categorized by segment quality scores to allow for TTS model training and fine-tuning. VALL-E and NaturalSpeech 2 systems are trained and fine-tuned on these subsets to validate the usability of WenetSpeech4TTS, establishing baselines on benchmark for fair comparison of TTS systems. The corpus and corresponding benchmarks are publicly available on 
    huggingface\footnote{\href{https://huggingface.co/datasets/Wenetspeech4TTS/WenetSpeech4TTS}{Huggingface link: https://huggingface.co/datasets/Wenetspeech4\\TTS/WenetSpeech4TTS}}.
    
\end{abstract}




\section{Introduction}

Recently, language model-based and diffusion-based \cite{ddpm} text-to-speech (TTS) approaches have demonstrated remarkable performance. 
In addition to the strong modeling capabilities of the models, these achievements are mainly attributed to the scale-up of the data.
For example, VALL-E \cite{valle} scaled up its training data to $60,000$ hours of English speech\cite{librilight} and NaturalSpeech 2 \cite{shen2024naturalspeech} used $44,000$ hours of transcribed English speech data\cite{librispeech}.
However, these existing large-scale open-source datasets are in English or multilingual. In contrast, there is a noticeable lack of comparably extensive datasets for Chinese TTS applications. To the best of our knowledge, the largest open-source transcribed Chinese speech dataset for TTS currently is DIDISPEECH\cite{didispeech}, which contains nearly $800$ hours of reading-style speech. This is insufficient for training large TTS models due to its small data scale and low diversity. Furthermore, there is currently a lack of a publicly available benchmark of TTS models trained on large-scale Chinese corpora for fair comparison. 
Due to the difficulty and high cost of reproducing and training other systems, some studies can only obtain samples for comparison and evaluation by downloading audio samples from the demo pages of other systems.
It poses a certain obstacle to the development of the community.

To the best of our knowledge, WenetSpeech\cite{wenetspeech} is the currently largest widely-used open-source Mandarin speech corpus for automatic speech recognition (ASR), containing $12,483$ hours of transcribed Mandarin speech data collected from \textit{YouTube} and \textit{Podcast}. The speech data in WenetSpeech is sourced from real recordings, covering a large number of speakers and a wide range of domains such as audiobooks, interviews, reading, etc.
Despite that some TTS studies \cite{valle-x} have used WenetSpeech as part of their training data to train large-scale models, the dataset itself exhibits certain limitations:
Most notably, the speech data, sourced directly from \textit{YouTube} and \textit{podcast}, contains various noises and distortions and lacks further post-processing and refinement. While such a complex acoustic environment might be beneficial for constructing robust ASR systems, it is definitely unsuitable for TTS systems.
Secondly, the segmentation of the data from \textit{YouTube} videos is performed by an optical character recognition (OCR) system. A subtitle phrase from the source video is often split into several pieces that appear across different frames by annotators. 
Despite efforts to merge phrases during the OCR segmentation process, this approach frequently leads to sentences being split into multiple speech segments, many of which are too short and semantically incomplete.
Furthermore, since it is specifically designed for ASR purposes, WenetSpeech does not ensure \textit{speaker homogeneity} within segments, leading to segments where multiple speakers are present.
Additionally, the speech and subtitle annotations of the original video data are not completely synchronized in time, leading to inaccurate subtitle boundaries detected by the OCR system. Consequently, some segments may have truncated words at the beginning or end.


To address the aforementioned issues of original WenetSpeech, in this paper, we introduce WenetSpeech4TTS, a $12,800$-hour large-scale dataset tailored for the training of speech synthesis models. Based on the speech quality, WenetSpeech4TTS provides three subsets: \emph{Basic, Standard, and Premium}, 
containing $7,226$, $4,056$, and $945$ hours of effective data, respectively. 

From WenetSpeech to WenetSpeech4TTS, we designed an automatic pipeline containing multiple processing steps, a similar approach of deriving TTS dataset from wild data have been proved effective in ~\cite{autoprep}\footnote{However, the processed dataset has not been released}. Specifically, we refined WenetSpeech for TTS by merging segments based on speaker similarity and pause duration, and by expanding segment boundaries to prevent truncated words. Audio quality was enhanced using the denoising model, followed by quality scoring. Furthermore, a speaker diarization system clustered segments from the same speaker, while a more advanced ASR system provided more accurate transcriptions.
To validate the usability of WenetSpeech4TTS, we trained the representative language model-based VALL-E\cite{valle} system and diffusion-based NaturalSpeech 2\cite{shen2024naturalspeech} system on different subsets. Both subjective and objective evaluations were conducted, demonstrating that WenetSpeech4TTS is effective for training large TTS models, with higher-quality subsets yielding better performance. 
The WenetSpeech4TTS corpus with all metadata and the corresponding benchmark are publicly available on huggingface.




\section{Processing pipeline}

This section will detail the processing pipeline that customizes the Wenetspeech4TTS dataset through a series of operations, mainly including adjacent segments merging, boundary extension, speech enhancement, multi-speaker detection, speech recognition, and quality filtering. 


\subsection{Adjacent segments merging}
\label{sec:asm}
\begin{figure}[th]
    \centering
    \includegraphics[width=0.9\linewidth]{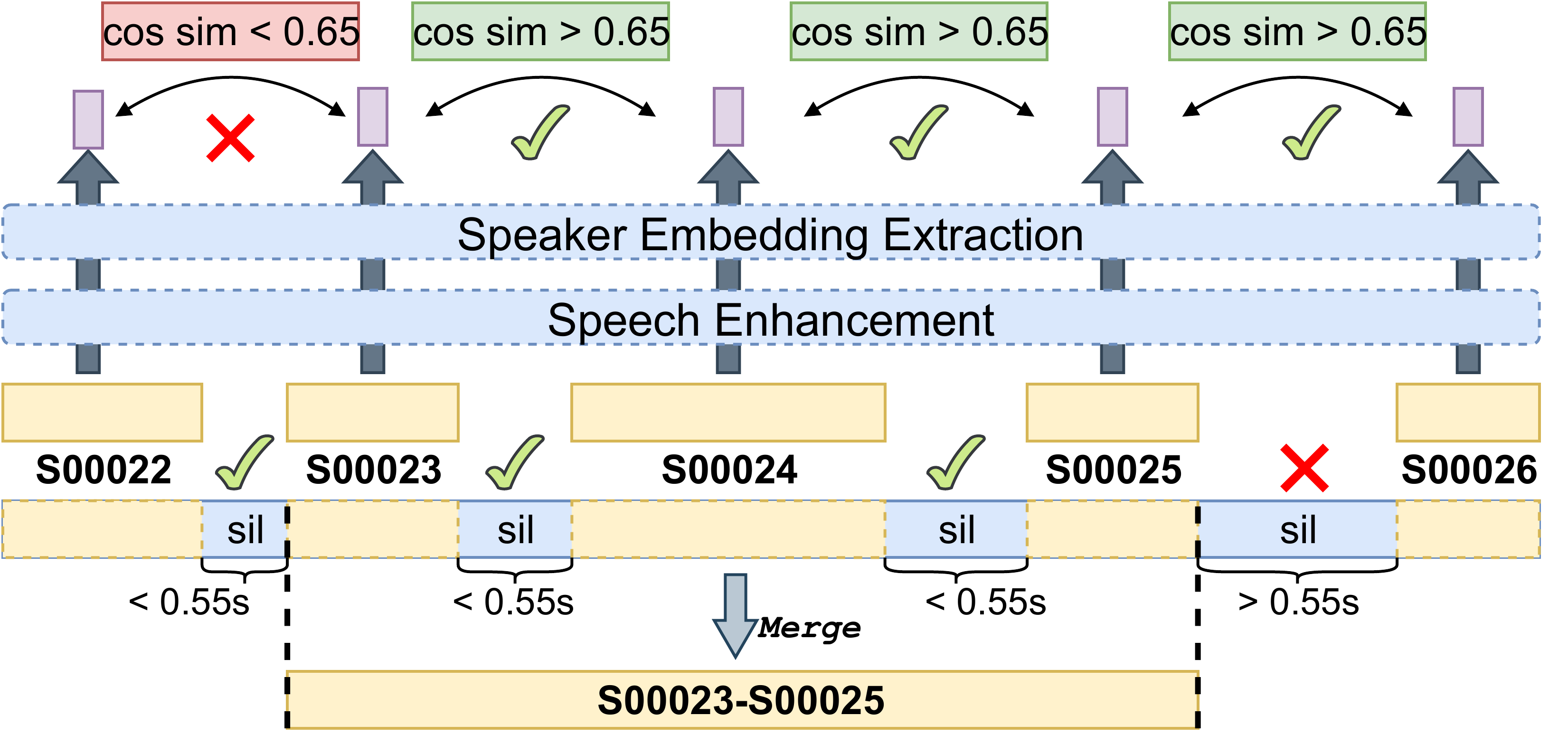}
    \caption{Demonstration of adjacent segments merging}
    \label{merge}
\end{figure}

To alleviate the issue that many segments are too short and semantically incomplete, we designed a merging strategy based on the interval time and speaker similarity. As illustrated in Figure~\ref{merge}, we first examine the interval time between adjacent segments based on the timestamp information in WenetSpeech. If the interval time is below $0.55$ seconds, we consider them potentially belonging to the same sentence. We then apply speech enhancement to both segments to improve the speech quality, the specifics of the speech enhancement model will be elaborated in Section~\ref{Subsection: SE}. For the extraction of speaker embeddings, Resemblyzer~\footnote{\hyperlink{https://github.com/resemble-ai/Resemblyzer}{https://github.com/resemble-ai/Resemblyzer}} is employed.
The cosine similarity between the two embeddings is then calculated as the speaker similarity. Segments with similarity scores above $0.65$ are presumed to be from the same speaker and are merged, this process is repeated until the sentence duration reaches $20$ seconds.

\vspace{-10pt}

\subsection{Boundary extension}
\begin{figure}[th]
    \centering
    \includegraphics[width=0.9\linewidth]{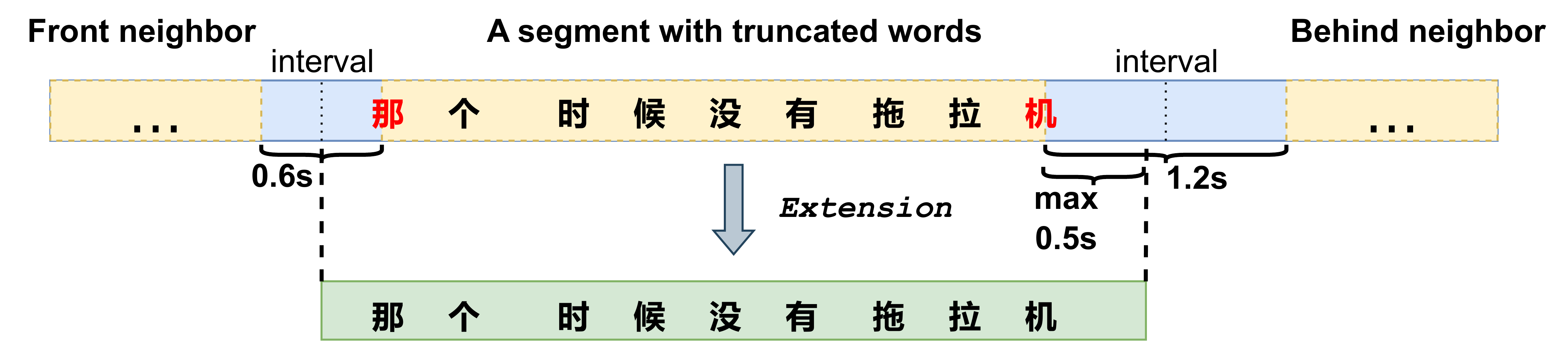}
    \caption{Boundary extension}
    \label{extension}
   
\end{figure}

To address the issue of truncated words at the beginnings and ends of segments, we proposed to extend the speech segments after the merging operation described in Section~\ref{sec:asm}. As illustrated in Figure~\ref{extension}, we extend the segment start backwards, without surpassing the midpoint of the interval with the previous neighbor, and the maximum extension is limited to $0.5$ seconds.
The same operation is also applied to the segment ends.

\vspace{-10pt}
\subsection{Speech enhancement}
\label{Subsection: SE}
\begin{figure}[htbp]
    \centering
    \includegraphics[width=0.9\linewidth]{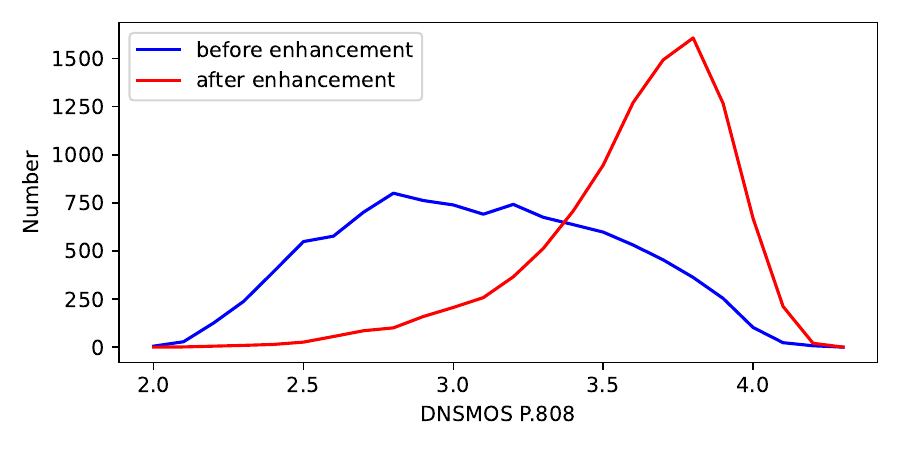}
    \caption{Distribution of DNSMOS P.808 scores for 10,000 random segments: enhanced (Red) vs. original (Blue).}
    \label{enhance}
\end{figure}

We refine the timestamp information of WenetSpeech based on the results of the previous two steps and then segment the source data to obtain new speech segments. The speech enhancement model was employed again to improve the speech quality of these segments. Specifically, we utilized MBTFNet\cite{MBTFnet}, a multi-band time-frequency neural network designed specially to remove noise, background music, or other forms of interference, to obtain clean vocals.
This speech enhancement operation removes interfering factors for subsequent operations such as quality evaluation, multi-speaker detection, speech recognition, and quality filtering.
To validate the efficacy of our speech enhancement model, we conducted a quality assessment on a random sample of $10,000$ segments after merging and extension, using the DNSMOS metric\cite{dnsmos2021, dnsmos2022p835}. We obtained and compared the DNSMOS P.808 scores for these segments before and after enhancement. The results, illustrated in Figure~\ref{enhance}, clearly demonstrate a significant improvement in segment quality following enhancement.

\subsection{Multi-speaker detection}

For TTS models, speech segments containing multiple speakers are often not directly usable for training. To ensure such ``speaker homogeneity'',
we adopt a clustering based diarization process to obtain speaker assignments on segments.
Specially, we train the ResNet293~\cite{wespeaker} speaker embedding model on VoxCeleb~\cite{voxceleb} and VoxCeleb2~\cite{voxceleb2} datasets.
Then we use the speech enhancement model in Section~\ref{Subsection: SE} to reconstruct the training data to fine-tune the speaker embedding model, eliminating the possible data mismatch introduced by the speech enhancement process.
Each segment is divide into chunks with a $1.5$s window and a $0.75$s shift to compute speaker embeddings. Subsequently, a standard spectral clustering algorithm is applied to obtain the cluster centers and assignments.  
We label the $10\%$ of embeddings in each cluster that are farthest from the cluster center as centrifugal embeddings.
To avoid situations where the same person is assigned to different clusters, we further merge the clusters if the cosine similarity between their cluster centers is greater than $0.75$.
Finally, if the speaker embeddings are assigned to different clusters or if more than half of the speaker embeddings are labeled as centrifugal in a segment, then the segment will be considered to contain more than one speaker.


\subsection{Speech recognition}
\label{recognition}
Compared to the original WenetSpeech, certain segments need to be transcribed since they have been restructured. Additionally, the transcripts obtained from the OCR system exhibit deletion errors in colloquial and repeated words, and the performance of the ASR system originally applied to the \textit{Podcast} data is outdated. 
Therefore, we use the open-source industrial-level Paraformer-large\cite{gao2022paraformer} system~\footnote{\hyperlink{https://www.modelscope.cn/models/iic/speech_paraformer-large-vad-punc_asr_nat-zh-cn-16k-common-vocab8404-pytorch/summary}{https://www.modelscope.cn/models/iic/speech\_paraformer-large-vad-punc\_asr\_nat-zh-cn-16k-common-vocab8404-pytorch/summary}} to obtain new text transcriptions for the speech segments, this model achieved a $6.74\%$ character error rate (CER) on the WenetSpeech ``Test\_Net'' set.
Meanwhile, we calculate the posterior probability of the Paraformer recognizing each segment as a confidence score.

\subsection{Quality filtering}
After the preceding steps, we conduct a quality-based data filtration. Segments with multiple speakers, as well as those with speech recognition confidence scores below $0.7$, are excluded. The refined dataset, designated as the WenetSpeech4TTS corpus, comprises $12,800$ hours of speech segments.
\textit{Some intervals between segments, which are not included in WenetSpeech, are merged into WenetSpeech4TTS through the merging operation. This enables us to obtain a larger dataset.}

\captionsetup[figure]{skip=2pt}

\section{The WenetSpeech4TTS corpus}
\label{corpus}

\begin{figure}[htbp]
    \centering
    \includegraphics[width=0.9\linewidth]{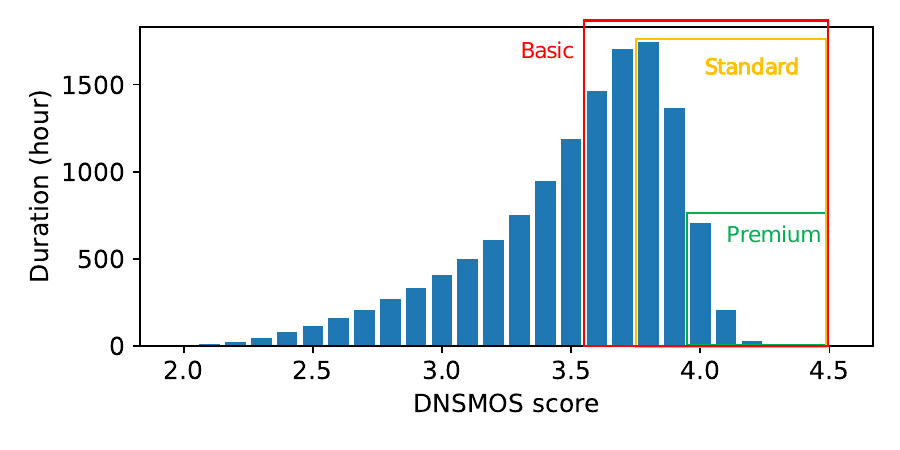}
    \caption{The distribution of speech data quality. The horizontal axis represents DNSMOS P.808 scores, and the vertical axis represents the scale of data corresponding to the scores.}
    \label{dur_mos}
\end{figure}

We further divide our WenetSpeech4TTS corpus into subsets based on the DNSMOS P.808 scores. 
Figure~\ref{dur_mos} illustrates the score distribution of the dataset.
Notably, segments with scores above $4.0$ are labeled as \textit{Premium}, those above $3.8$ as \textit{Standard}, and those above $3.6$ as \textit{Basic}. The remaining segments with a score lower than $3.6$ are labeled as \textit{Rest}.
Specifics about these subsets are shown in Table~\ref{subsets}.


\captionsetup[table]{skip=2pt}

\begin{table}[htbp]
\caption{The training data subsets.}
\label{subsets}
\resizebox{\columnwidth}{!}{%
\begin{tabular}{cccc}
\toprule[\heavyrulewidth]
\begin{tabular}[c]{@{}c@{}}Training\\ Subsets\end{tabular} & \begin{tabular}[c]{@{}c@{}}DNSMOS \\ Threshold\end{tabular} & Hours & \begin{tabular}[c]{@{}c@{}}Average Segment\\ Duration (s)\end{tabular} \\ \midrule
\textit{Premium}      & 4.0       & 945  & 8.3                 \\
\textit{Standard}        & 3.8      & 4,056 & 7.5                \\
\textit{Basic}           & 3.6          & 7,226 & 6.6                  \\ 
\textit{Rest}           & $<$ 3.6          & 5,574 & -                  \\ \hline
WenetSpeech4TTS (sum)    & -    & 12,800    & -    \\ \hline
WenetSpeech (orig)    & -         & 12,483 & -          \\ 
\bottomrule[\heavyrulewidth]
\end{tabular}%
}
\end{table}

\begin{figure}[htbp]
\vspace{-10pt}
    \centering
    \includegraphics[width=0.9\linewidth]{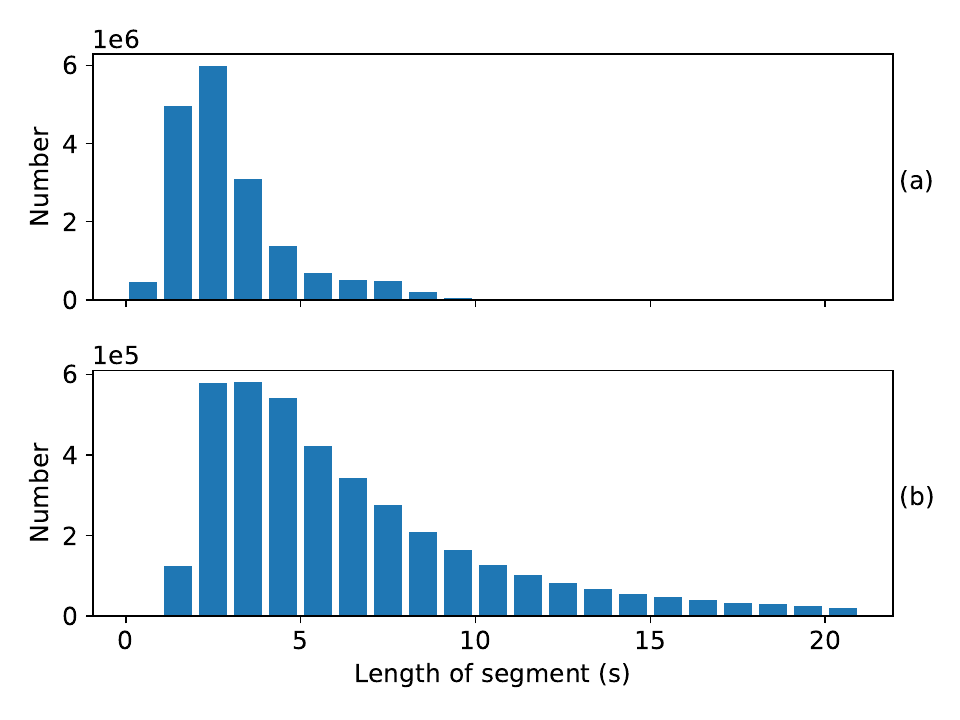}
    \caption{The distribution of segment lengths in WenetSpeech (a) and the WenetSpeech4TTS \textit{Basic} subset (b).}
    \label{subplot}
\end{figure}

Figure~\ref{subplot} shows the distribution of segment lengths in Wenetspeech and the distribution of segment lengths in the WenetSpeech4TTS \textit{Basic} subset.
After the merging operation, the number of segments under $3$ seconds greatly decreased, while segments over $5$ seconds significantly increased. Importantly, many previously rare segments over $10$ seconds are generated. This benefits training TTS models and demonstrates that the merging strategy is reasonable and necessary.

The WenetSpeech4TTS test set comprises $150$ test sentences and $26$ target speakers, of which $10$ in-set speakers from the WenetSpeech4TTS corpus and $16$ out-of-set speakers, with equal numbers of males and females. The out-of-set speakers include $6$ from WenetSpeech Test\_Net set, $5$ from the open-source dataset Aishell3~\cite{aishell3}, and the other $5$ are amateur speakers collected.

The WenetSpeech4TTS corpus, including segments, transcripts, and DNSMOS scores, is open-sourced and audio samples can be available at our demo page~\footnote{\hyperlink{https://wenetspeech4tts.github.io/wenetspeech4tts/}{https://wenetspeech4tts.github.io/wenetspeech4tts/}}.

\begin{table*}

\centering
\caption{The results of VALL-E and NaturalSpeech 2 with $95\%$ confidence intervals. The training set \textit{Basic} means training with the \textit{Basic} subset. \textit{Standard},FT means fine-tuning with the \textit{Standard} subset. \textit{Premium},FT means fine-tuning with the \textit{Premium} subset after \textit{Standard},FT.}
\label{results}
\resizebox{0.86\textwidth}{!}{%
\begin{tabular}{cccccc|cccc}
\toprule[\heavyrulewidth]
\multirow{2}{*}{Method} & \multirow{2}{*}{\begin{tabular}[c]{@{}c@{}}Training\\ set\end{tabular}} & CER $\downarrow$  & SECS $\uparrow$ & NMOS $\uparrow$     & SMOS $\uparrow$      & CER $\downarrow$   & SECS $\uparrow$  & NMOS $\uparrow$     & SMOS $\uparrow$    \\ \cline{3-10} 
                        &                                & \multicolumn{4}{c|}{$10$ seen speakers} & \multicolumn{4}{c}{$16$ unseen speakers} \\ \midrule[\heavyrulewidth]
\multirow{3}{*}{VALL-E} & \textit{Basic}                              & $13.12$  & $0.824$ & $3.34\pm0.15$ & $\mathbf{3.68\pm0.12}$ & $14.96$  & $\mathbf{0.781}$ & $3.27\pm0.12$ & $3.37\pm0.12$ \\
                        & \textit{Standard}, FT                           & $10.23$  & $0.811$ & $3.41\pm0.14$         & $3.59\pm0.13$         & $12.91$  & $0.772$ & $3.36\pm0.10$         & $3.46\pm0.11$         \\
                        & \textit{Premium}, FT                           & $\mathbf{8.10}$   & $\mathbf{0.827}$ & $\mathbf{3.50\pm0.15}$ & $3.61\pm0.13$ & $\mathbf{9.82}$  & $\mathbf{0.781}$ & $\mathbf{3.45\pm0.09}$ & $\mathbf{3.58\pm0.09}$         \\ \hline
\multirow{4}{*}{NS2}    & \textit{Basic}                              & $11.88$ & $0.783$ & $2.94\pm0.13$         & $3.15\pm0.12$         & $14.06$  & $0.756$ & $2.64\pm0.11$         & $2.95\pm0.11$         \\
                        & \textit{Standard}, FT                           & $8.28$  & $0.803$ & $3.38\pm0.14$         & $\mathbf{3.50\pm0.13}$         & $10.78$  & $0.766$ & $3.14\pm0.10$         & $3.29\pm0.10$         \\
                        & \textit{Premium}, FT                           & $\mathbf{6.30}$  & $\mathbf{0.820}$  & $\mathbf{3.40\pm0.13}$         & $3.48\pm0.13$         & $\mathbf{7.89}$   & $\mathbf{0.783}$ & $\mathbf{3.36\pm0.09}$   & $\mathbf{3.45\pm0.10}$         \\
                                  \bottomrule[\heavyrulewidth]
\end{tabular}%
}
\end{table*}

%
%
%
%
%
%

\section{Experiments}

To validate the usability of the WenetSpeech4TTS dataset and establish baselines on benchmark for large-scale TTS models, two popular large-scale TTS models, VALL-E~\cite{valle} and NaturalSpeech 2~\cite{shen2024naturalspeech}, are trained on the \textit{Basic} set and finetuned on the other two sets with higher quality successively. We open-source these model weights along with our dataset.

\subsection{VALL-E and NaturalSpeech 2}

In VALL-E, both the Autoregressive (AR) and Non-Autoregressive (NAR) have the same transformer architecture with $12$ layers, $16$ heads, an embedding dimension of $1024$, a feed-ward layer dimension of $4096$, and a drop out of $0.1$. 
To get better speech quality, we employ AudioDec~\footnote{\hyperlink{https://github.com/facebookresearch/AudioDec}{https://github.com/facebookresearch/AudioDec}}as an acoustic codec rather than Encodec~\cite{10096509,defossez2022highfi}. We resample all recordings to $24$k Hz and use the \textit{libritts v1} model to extract acoustic tokens. For all transcripts, we use the front end of BERT-VITS2~\footnote{\hyperlink{https://github.com/fishaudio/Bert-VITS2}{https://github.com/fishaudio/Bert-VITS2}} to convert text to Pinyin. Both the AR model and NAR model are trained using $4$ NVIDIA A40 GPUs for a week, respectively. Initially, we use WenetSpeech4TTS \textit{Basic} as the training set for the foundational VALL-E model. To improve audio quality and performance, we fine-tune the foundational VALL-E model with WenetSpeech4TTS \textit{Standard} and then with WenetSpeech4TTS \textit{Premium}.



The model configurations of NaturalSpeech 2 follow the default configurations in the Amphion toolkit\footnote{\hyperlink{https://github.com/open-mmlab/Amphion}{https://github.com/open-mmlab/Amphion}} ~\cite{zhang2023amphion}.
The data preprocessing process is consistent with VALL-E, except that we use the default Encodec as the acoustic codec of NaturalSpeech 2.
Additionally, we sample $2,000$ hours of labeled data from WenetSpeech4TTS \textit{Basic} to train the ASR model in Kaldi~\footnote{\hyperlink{https://github.com/kaldi-asr/kaldi}{https://github.com/kaldi-asr/kaldi}} toolkit, and then obtain external duration information of all labeled data for NaturalSpeech 2 training. 
We train the model for a total of $600$k steps on $8$ NVIDIA A6000 GPUs. 
The training and fine-tuning strategy of NaturalSpeech 2 is similar to that of VALL-E. 

\subsection{Evaluation metrics}
We use the test set introduced in Section~\ref{corpus} to test these models' performance.
We synthesized all 150 utterances for each speaker for objective evaluation.
We randomly selected 5 seen speakers and 5 unseen speakers, with 8 random samples for each speaker, for a total of 80 test cases for subjective evaluation.
The objective evaluations included speaker embedding cosine similarity (SECS) and character error rate (CER) in ASR. The SECS metric was computed by extracting speaker embeddings with Resemblyzer~\footnote{\hyperlink{https://github.com/resemble-ai/Resemblyzer}{https://github.com/resemble-ai/Resemblyzer}} and calculating the cosine similarity. 
The CER was measured between transcripts of real and synthesized utterances transcribed by Paraformer employed in Section~\ref{recognition}.
For subjective evaluation, We conduct Mean Opinion Score (MOS) experiments to evaluate speech naturalness (NMOS) and speaker similarity MOS (SMOS) reference to target speaker utterances. In each MOS test, a group of $20$ native Chinese Mandarin listeners are asked to listen to synthetic speech and rate on a scale from $1$ to $5$ with a $0.5$-point interval. 
The objective and subjective results are shown in Table~\ref{results}.

\subsection{Objective evaluation}
For all models, the CER and SECS metrics are worse for unseen speakers compared to seen speakers, indicating that all models obtain better intelligibility and similarity on speakers they have seen during training. 
The CER for both VALL-E and NaturalSpeech 2 significantly decreases as the quality of the training data improves (from \textit{Basic} to \textit{Premium}, FT), which suggests that higher-quality subsets lead to more stable speech synthesis. 
NaturalSpeech 2 outperforms VALL-E in terms of CER across all training sets as well as all speakers. This indicates that the NaturalSpeech 2 approach, which benefits from its external duration-aligned training, is more robust and achieves better intelligibility than VALL-E's autoregressive modeling.
Across both seen and unseen speakers, the SECS metric is relatively high for all models, indicating that their ability to model speaker timbre was generally good.
As the quality of training data improves, the SECS metric of VALL-E remains relatively stable while that of NaturalSpeech 2 shows an upward trend. This suggests that NaturalSpeech 2 is more affected by data quality in speaker timbre modeling.

\subsection{Subjective evaluation}
The SMOS and NMOS scores of both VALL-E and NaturalSpeech 2 for seen speakers are higher than those for unseen speakers, which is the same situation as objective evaluation.
As data quality improves, NMOS scores generally show an upward trend. Except that SMOS scores of VALL-E on in-set speakers are relatively stable, the other SMOS scores also gradually improve with the improvement of data quality.
Overall, the naturalness of VALL-E is better than that of NaturalSpeech 2 under the same conditions. We attribute the reason to the error accumulation of NaturalSpeech 2's external duration alignment model using automatically transcribed data. 
In the objective evaluation, the speaker similarity of VALL-E and NaturalSpeech 2 is basically the same, while in the subjective evaluation, the speaker similarity of NaturalSpeech 2 is lower. We speculate that it may be because the speech quality generated by Encodec is worse than that of AudioDec, which affects the listener's perception of speaker timbre.
Audio samples can be available at our demo page~\footnote{\hyperlink{https://wenetspeech4tts.github.io/wenetspeech4tts/}{https://wenetspeech4tts.github.io/wenetspeech4tts/}}.

\section{Conclusion}
This paper proposed \textit{WenetSpeech4TTS}, a multi-domain Mandarin corpus for large TTS model training which is derived from the open-source WenetSpeech dataset.
We designed a series of processing operations to refine the WenetSpeech data and then divided it into subsets of different sizes according to the quality of the speech.
To prove the usability of WenetSpeech4TTS and provide a benchmark, we trained and fine-tuned VALLE and NaturalSpeech 2 on these subsets. Experimental results demonstrate that WenetSpeech4TTS can be used to train large TTS models and that higher-quality subsets achieve better performance.

\bibliographystyle{IEEEtran}
\bibliography{mybib}

\end{document}